# Operating Principles of Vertical Transistors Based on Monolayer Two-Dimensional Semiconductor Heterojunctions

Kai Tak Lam, Gyungseon Seol and Jing Guo

Department of Electrical and Computer Engineering, University of Florida, Gainesville, FL 32611-6130

ABSTRACT

A vertical transistor based on a double gated, atomically thin heterojunction is theoretically examined. Both p-type and n-type transistor operations can be conveniently achieved by using one of the two gates as the switching gate. The transistor shows excellent saturation of output I-V characteristics due to drain-induced depletion and lack of tunneling barrier layers. The subthreshold slope could be below the thermionic limit due to band filtering as the switching mechanism. The atomically thin vertical PN heterojunction can be electrostatically modulated from a type II heterojunction to a broken bandgap alignment, which is preferred for maximizing the on-current.

Two-dimensional (2D) layered dichalcogenide materials, which offer a bandgap that is absent in graphene, have attracted extensive research interest for potential electronic device applications[1]. Vertical tunneling field effect transistors (FETs), which uses graphene as the source and drain contacts, and sandwiched BN or 2D dichacolgenide layers as the tunneling barrier, have been experimentally fabricated and theoretically examined [2-6]. Several orders of magnitude in on-off current ratio have been achieved. The on-current, however, is limited by the tunneling barrier and the output I-V characteristic lacks a saturation region [2], which is a necessary requirement for many digital and analog circuit applications. On the other hand, atomically thin PN heterojunctions between monolayer dichacolgenide materials have been experimentally demonstrated, which show rectifying I-V characteristics and photocurrent modulated by a gate bias [7]. Recently, two-dimensional heterojunction interlayer tunneling FETs have been theoretically examined [8], and a study on their transfer characteristics suggested that this tunneling FETs could achieve ultrasteep subthreshold slope.

In this work, we theoretically model a vertical FET formed with a heterojunction between a monolayer N type and P type 2D material modulated by double gates. The output characteristics of the heterojunction FETs show excellent saturation due to drain induced depletion. Furthermore, without any tunneling barrier layer between the N type and P type materials, the transistor is expected to deliver a larger on-current compared to those with tunneling barrier layers. Both P type and N type operations of the transistor, which are crucial for potential CMOS and analogue applications of the device, can be achieved by properly choosing a switching gate. Electrostatic modulation of atomically thin PN heterojunctions is also studied.

The modeled device structure is shown in Fig. 1. The atomically thin heterojunction is formed between $WTe_2$ and $MoS_2$ monolayers, which are modulated by top and bottom gates with a gate insulator thickness of $t_{ox}$=3nm and dielectric constant of $\kappa$=20. The affinity and bandgap of monolayer $MX_2$ dichacolgenide materials have been studied before. As shown in Fig. 1(b), the band alignment in the absence of gating is determined according to the *ab intio* simulation results reported in Ref. [9], and the qualitative conclusions are insensitive to the uncertainty of the equilibrium band alignment. We focus our attention on the intrinsic properties of the vertical field-effect transistor (FET) formed by the double gated monolayer heterojunction, and assume that the $MoS_2$ and $WTe_2$ out of the gating region are sufficiently heavily doped [10, 11], so that the extension regions do not play a role in intrinsic vertical transistor switching.

Self-consistent electrostatics plays an important role in transistor characteristics. In this work, the carrier statistics equations in $WTe_2$ and $MoS_2$ monolayers with Fermi levels split by the applied voltage are solved self-consistently with Poisson equation in the form of a capacitance model. The approach of calculating the self-consistent electrostatic potential is same as what was described before in vertical tunneling transistors [5]. After the self-consistent electrostatic potential is obtained, the source-drain current is computed as the interlayer current by using the Landauer-Buttiker formula [12],

$$I = \frac{g_s e}{h} \sum_{k_t, k_b} \int dE \cdot T_{tb}(E)[f_t(E) - f_b(E)], \tag{1}$$

where $g_s$ is the spin degeneracy factor, $f_{t,b}(E)$ is the Fermi Dirac distribution function, and $T_{tb}(E)$ is the interlayer transmission between the wave state with a wave vectors of $\vec{k}_t$ in the top layer and that in the bottom layer with a wave vector of $\vec{k}_b$. The interlayer transmission can be computed as follows if the interlayer coupling is weak compared to intralayer binding [13],

$$T_{tb}(E) = A_t(E) M_{tb} A_b(E) M_{tb}^\dagger, \tag{2}$$

where $M_{tb}$ is the matrix element of the scattering potential between the wave states. The spectral function is

$$A_{t,b}(E) = 2\pi\delta[E - E_{t,b}(\vec{k}_{t,b})], \tag{3}$$

where $E_{t,b}(\vec{k}_{t,b})$ is the E-k relation of the top and bottom layer. Integrating Eq. (1) over energy results in the following expression [8, 14],

$$I = \frac{2\pi g_s e}{\hbar}\sum_{k_t,k_b}|M_{tb}|^2[f_t(E_t) - f_b(E_b)]\delta(E_t - E_b), \tag{4}$$

and the current density is

$$J = \frac{2\pi g_s e}{\hbar}\int\frac{dk_t^2}{(2\pi)^2}\int\frac{dk_b^2}{(2\pi)^2}|M_{tb}|^2 S[f_t(E_t) - f_b(E_b)]\delta(E_t - E_b), \tag{5}$$

where $S$ is the area of the monolayer heterojunction. The following argument can be made on why the current density $J$ is independent of area $S$. With one interlayer scatterer, the potential matrix element scales as $1/S$. In the presence of $N$ uncorrelated scatterers, $|M_{tb}|^2 S$ scales as $\frac{N}{S} = n_I$, where $n_I$ is the area density of interlayer scatterers. $|M_{tb}|^2 S$, therefore, is independent of the area $S$ and is proportional to $n_I$.

Interlayer scattering events play an important role in conserving momentum and energy in interlayer transport. The exact mechanisms of interlayer scattering are unclear, which could be sample-dependent and due to charge impurities, disorders and etc. The uncertainty hinders first principle calculation of the matrix element $M_{tb}$, which can be a fitting parameter from experimental data. Approximations, however, can be made on the matrix element to facilitate study of the device characteristics. If the in-plane component of the interlayer scattering potential is a short range potential and can be approximated by a delta function, its Fourier component is wave vector independent, $|M_{tb}|^2 S = M_0^2 S$. In this case, Eq. (5) can be simplified to a form proportional to the product of density of states (DOS) [2, 3],

$$J = \frac{2\pi g_s e}{\hbar} \int dE \cdot D_t(E) D_b(E) |M_0|^2 S [f_t(E) - f_b(E)], \qquad (6)$$

where $D_{t,b}(E)$ is the DOS of the top and bottom layer without spin degeneracy. The current density can also be expressed as,

$$J = g_s e \int dE \frac{D_{t(b)}(E)}{\tau_{t(b)}} [f_t(E) - f_b(E)], \qquad (7)$$

where the interlayer transport rates of the carriers $\tau_{t(b)}$ in the top (bottom) layer is determined by an expression like the Fermi's golden rule,

$$\frac{1}{\tau_{t(b)}} = \frac{2\pi}{\hbar} D_{b(t)}(E) M_0^2 S. \qquad (8)$$

Furthermore, to examine the effect of wavevector dependent interlayer scattering potential, the matrix element in the form of [8]

$$|M_{tb}|^2 S = |M_{00}|^2 \frac{\pi L_C^2}{(1+|\vec{q}|^2 L_C^2/2)^{3/2}}, \qquad (9)$$

is used in the calculation of current in Eqs. (1) and (2), where $M_{00}$ is a constant in the unit of eV, the wavevector $\vec{q} = \vec{k}_t - \vec{k}_b$, and $L_C$ is a parameter in the unit of length.

The spectral function in Eq. (3) is under the approximation of no broadening, which requires high quality of semiconductors and weak interlayer coupling. Introducing a finite broadening $\eta$ results in the following expression,

$$A_{t,b}(E) = 2 Im \left( \frac{1}{E - E_{t,b}(\vec{k}_{t,b}) + i\eta_{t,b}} \right), \qquad (10)$$

which leads to bandgap states and describes the finite carrier lifetime due to interlayer coupling and disorder scattering. The impact on I-V characteristics can be assessed by using Eq. (9) in the calculation of current.

In order to understand the switching mechanism of the vertical monolayer heterojunction FET, we first examine the band profiles at the off and on states, as shown in Fig. 2. The top gate voltage is fixed at $V_{TG}$=-0.5 V, which results in p-type electrostatic doping of the WTe$_2$ and the Fermi level of the top layer $E_{Ft}$ below the valence band of WTe$_2$. At a low bottom gate voltage of $V_{BG}$=0V, the carrier density in the MoS$_2$ layer is low, and the Fermi levels of both the top and bottom layers, $E_{Ft}$ and $E_{Fb}$, are in the bandgap of MoS$_2$, and the source-drain current is blocked by the MoS$_2$ bandgap. As the bottom gate voltage increases, the conduction band edge of MoS$_2$ moves below the valence band edge and the Fermi level of WTe$_2$, which results in an on state as shown in Fig. 2b. This band filtering mechanism is responsible for the switching of the vertical monolayer heterojunction FET.

The switching mechanism above shares similarity with the band-to-band tunneling FETs, which can offer ultrasteep subthreshold slope below the thermionic limit of 60mV/dec at room temperature[15]. Important difference, however, exists. Both the n-type and p-type semiconductor layers in the heterojuction FET are only monolayer thick, which enables effective electrostatic modulation of the heterojunction interface. The top and bottom gates directly modulate all atoms at the interface which provides advantages in terms of device operation. For example, the on-current of a band-to-band tunneling FET is often limited by band-to-band tunneling, and a broken bandgap heterojunction is preferred but difficult to achieve. For the vertical monolayer heterojunction FET, electrostatic modulation of the heterojunction interface can be so effective that the interface can be modulated from a type II heterojunction at the off state to a broken bandgap heterojunction at the on state, as shown in Fig. 2(b), which is preferred for maximizing the on-current.

The switching I-V characteristics of the vertical heterojunction FETs are shown in Fig. 3. By fixing the bottom gate voltage to electrostatically dope the $MoS_2$ layer to n-type, p-type transistor switching is achieved by using the top gate as the switching gate. On the other hand, by using the bottom gate as the switching gate, n-type switching is achieved. Both n-type and p-type transistor characteristics, which are crucial for CMOS applications, can be conveniently achieved by choosing one of the two gates as the switching gate. Figure 3 also shows that due to the switching mechanism of band filtering, the subthreshold slope could be very steep and is limited by bandgap states, as discussed before in 2D semiconductor interlayer tunneling FETs [8].

The minimal leakage current shows different trends for p-type operation and n-type operation due to the following reason. The minimal leakage current is limited by thermionic emission in the conduction band of $WTe_2$ and the valence band of $MoS_2$. Because of the much smaller bandgap of $WTe_2$, thermionic emission in the conduction band of $WTe_2$ dominates. For n-type transistor operation in Fig. 3(b), the $WTe_2$ layer, which is grounded, acts as the source, whose Fermi level remains relatively constant to its conduction band edge. As a result, the minimal leakage current is insensitive to the applied drain voltage on $MoS_2$. For p-type transistor operation, the drain voltage is applied on the $WTe_2$ layer and as the drain voltage increases, the $WTe_2$ Fermi level moves closer to the conduction band edge, which results in an increase of the minimal leakage current, as shown in Fig. 3(a).

Tunneling barrier material layers, such as layers of BN or $MoS_2$, were inserted between the source and drain graphene contact layers in previously studied tunneling FETs [2-6, 8]. However, the tunneling layer, which is necessary to reduce the leakage current when graphene or metal source and drain contacts are used, restricts the on-current performance. Furthermore, the output I-V characteristics do not saturate, which severely limits the potential application in digital and analog electronics. By using a gated semiconductor PN heterojunction, no tunneling barrier layer is needed in the modeled transistor, resulting in a larger on-current. Furthermore, the output I-V characteristics of the heterojunction FETs show excellent saturation behavior as shown in Fig. 4(a). The saturation of the drain current can be explained by drain-induced depletion of the drain monolayer. As shown in Fig. 4(b), when the drain voltage increases, its Fermi level moves into the bandgap region, and the drain layer becomes depleted. The charge density and band edge in the drain monolayer are insensitive to further increase of the drain voltage, and the source-drain current saturates.

The above I-V characteristics are calculated based on Eq. (6), without considering bandgap states. What if the interlayer scattering potential is wavevector dependent and bandgap states are induced by broadening? To include these effects, we compute the device I-V characteristics using Eq. (1), with the wavevector dependent matrix element described in Eq. (9) and broadened spectral function described in Eq. (10). While the qualitative conclusions of both n-type and p-type operations, as well as saturation of the output I-V characteristics, remain the same, the bandgap states due to broadening have a considerable effect on the off-current and the subthreshold slope. As shown in Fig. 5, the off-current increases proportionally with the broadening due to the bandgap states, and the current above the threshold is insensitive to a small value of broadening. The subthreshold slope could still be below the thermionic limit if the broadening is small. The leakage current, however, is limited by the bandgap states.

In summary, electrostatic modulability of a monolayer vertical heterojunction is advantageous for monolayer heterojunction transistors. Properly chosen 2D monolayer materials can form type II heterojunction band alignment, which blocks the source-drain current by the semiconductor bandgaps at the off state, and electrostatic gating could lead to broken bandgap alignment at the on state. The atomic layer structure ensures that all atoms at the interface are efficiently modulated by electrostatic gating. A double gate configuration allows both n-type and p-type operations by using one of the two gates as the switching gate. The output I-V characteristics saturate due to drain induced depletion of the monolayer.

The authors are indebted to Prof. P. Kim of Columbia University for extensive technical discussions. This works was supported by NSF.

FIGURES

Figure 1. (a) Modeled device structure. A double gated vertical monolayer WTe$_2$-MoS$_2$ heterojunction. The gate insulator has a thickness of $t_{ox}$=3nm and a dielectric constant of κ=20. The work function of both gates is assumed to be $\Phi = 4.3eV$. (b) The band alignment of a WTe$_2$-MoS$_2$ monolayer heterojunction in absence of gating.

Figure 2. The band profile at (a) the off state, $V_{BG} = 0V$, and (b) the on state $V_{BG} = 0.3V$. The top gate voltage is fixed at $V_{TG} = -0.5V$ and the bottom gate voltage is varied, which corresponds to n-type operation. The WTe$_2$ is grounded with $E_{Ft} = 0$ and the MoS$_2$ layer has an applied drain voltage $V_D$=0.3eV, which results in $E_{Fb} = -0.3eV$. The modeled device is shown in Fig. 1(a).

Figure 3. (a) $I_D$ vs. $V_{TG}$ characteristics at different $V_D$ (from -0.05V to -0.3V at -0.05V/step) when the top gate acts as the switching gate. The bottom gate voltage is fixed at $V_{BG} = 0.3V$. The transistor operates as a p-type FET, in which the source (MoS$_2$ layer) is grounded. (b) $I_D$ vs. $V_{BG}$ characteristics at different $V_D$ (from 0.05V to 0.3V at 0.05V/step) when the bottom gate acts as the switching gate. The top gate voltage is fixed at $V_{TG} = -0.5V$. The transistor operates as an n-type FET, in which the source (WTe$_2$ layer) is grounded. The current is computed using Eq. (6) with interlayer transport time $\tau_t = 10ps$, which results in a matrix element of $M_0^2 S \approx 5.0 meV^2 nm^2$.

Figure 4. *Saturation of output I-V characteristics*: (a) I$_D$ vs. V$_D$ characteristics at different bottom gate voltages $V_{BG}$ (from 0.05V to 0.3V at 0.05V/step) in n-type operation. The top gate voltage is fixed at $V_{TG} = -0.5V$. (b) Bandprofile at $V_D$= 0.3V. The top gate voltage is $V_{TG} = -0.5V$ and the bottom gate voltage is $V_{BG} = 0.3V$. As the drain voltage increases, the drain Fermi level

moves into the bandgap, which results in full depletion of the MoS$_2$ layer and saturation of the source-drain current. The current is computed using Eq. (6) with $\tau_t = 10ps$, which results in a matrix element $M_0^2 S \approx 5.0 meV^2 nm^2$.

Figure 5. I$_D$ vs.V$_{BG}$ characteristics at $T$=300K in the presence of broadening induced bandgap states and momentum-dependent interlayer scattering potential. The parameters used for the interlayer scattering potential is $|M_{00}| = 0.2 meV$ and $L_c = 20 nm$, and the broadening is $\eta = 0.1$ meV (blue solid), 1meV (red dotted) and 5meV (green dashed) for both the top and bottom layers. The reference line of 60mV/dec is also shown. The top gate voltage is fixed at $V_{TG} = -0.5V$, and the drain voltage is $V_D$=0.3V.

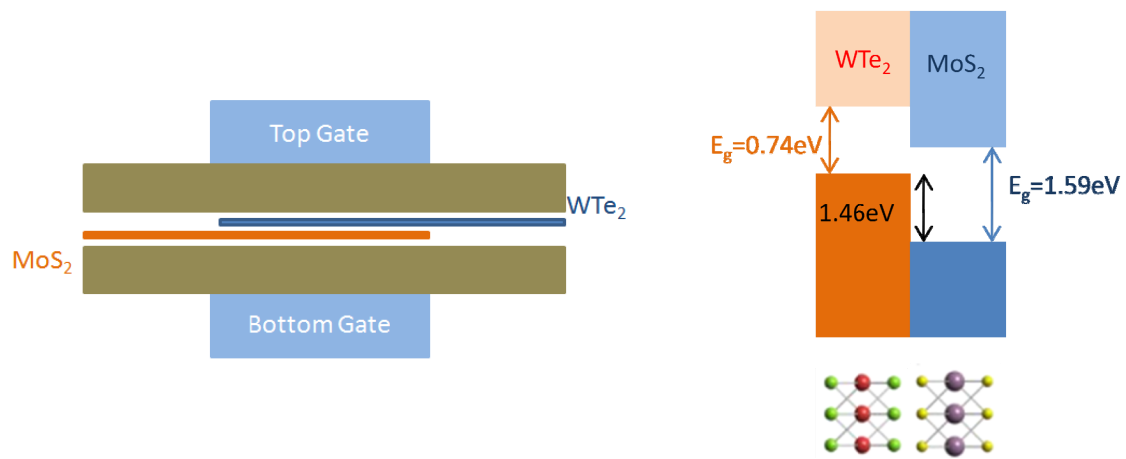

Figure 1

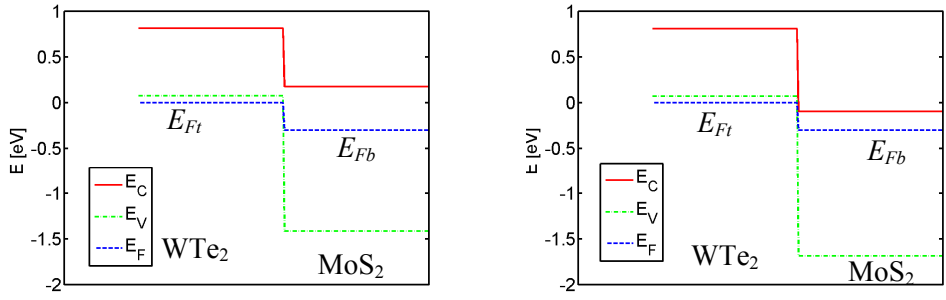

Figure 2

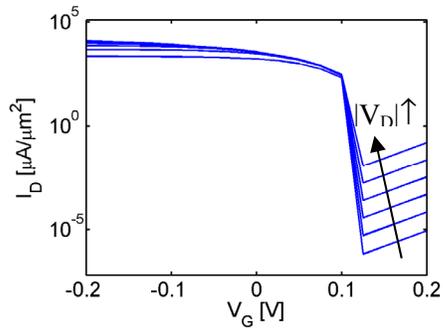 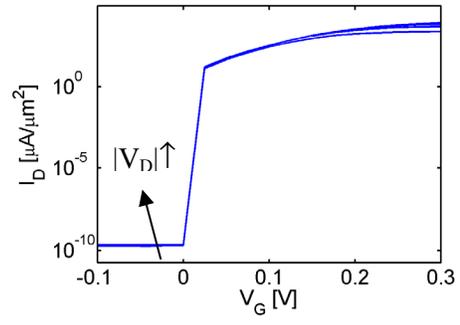

Figure 3

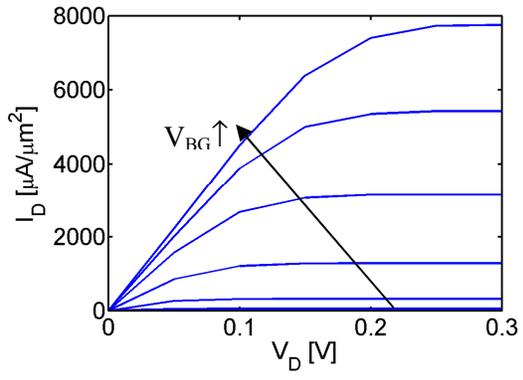 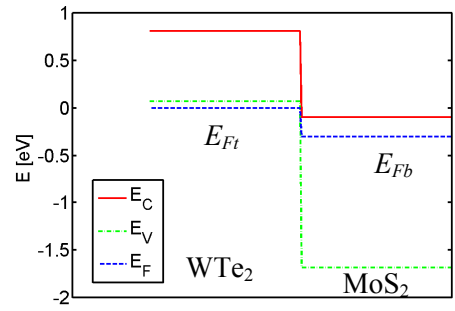

Figure 4

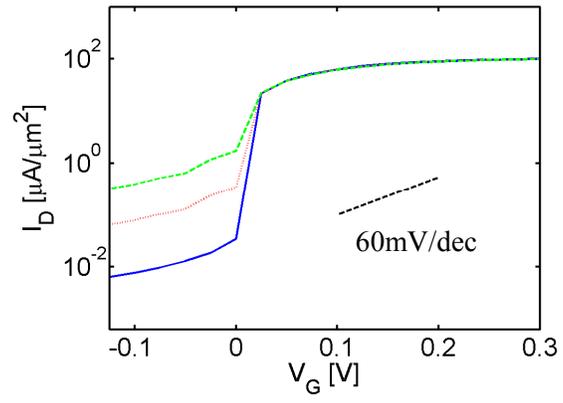

Figure 5.